\documentclass[amsmath, amssymb,floatfix,twocolumn,superscriptaddress,showpacs,showkeys,reprint,prb]{revtex4-1}
\usepackage{graphicx}
\usepackage{color}
\usepackage{verbatim}
\usepackage{textcomp}
\usepackage{subfigure}
\usepackage{mathbbol}
\usepackage[utf8]{inputenc}
\usepackage{float}
\usepackage{hyperref}
\usepackage{circuitikz}
\usepackage{tikz}
\usepackage{pgfplots}
\usepackage{amsmath}

\pgfplotsset{compat=newest}
\usepgfplotslibrary{external} 
\usetikzlibrary{plotmarks}

\newlength\fheight 
    \newlength\fwidth

\begin{document}
    \setlength\fheight{0.55\columnwidth} 
   \setlength\fwidth{0.7\columnwidth}

\title{One-dimensional Josephson junction arrays: Lifting the Coulomb blockade by depinning}
\author{Nicolas \surname{Vogt}}
\affiliation{Institut f\"ur Theorie der Kondensierten Materie,  
Karlsruhe Institute of Technology, D-76128 Karlsruhe, Germany}
\author{Roland \surname{Sch\"afer}}
\affiliation{Institut f\"ur Festk\"orperphysik,  
Karlsruhe Institute of Technology, D-76021 Karlsruhe, Germany}
\author{Hannes \surname{Rotzinger}}
\affiliation{Physikalisches Institut,  
Karlsruhe Institute of Technology, D-76128 Karlsruhe, Germany}
\author{Wanyin \surname{Cui}}
\affiliation{Physikalisches Institut,  
Karlsruhe Institute of Technology, D-76128 Karlsruhe, Germany}
\affiliation{Institut f\"ur Festk\"orperphysik,  
Karlsruhe Institute of Technology, D-76021 Karlsruhe, Germany}
\author{Andreas \surname{Fiebig}}
\affiliation{Physikalisches Institut,  
Karlsruhe Institute of Technology, D-76128 Karlsruhe, Germany}
\affiliation{Institut f\"ur Festk\"orperphysik,  
Karlsruhe Institute of Technology, D-76021 Karlsruhe, Germany}
\author{Alexander \surname{Shnirman}}
\affiliation{Institut f\"ur Theorie der Kondensierten Materie,  
Karlsruhe Institute of Technology, D-76128 Karlsruhe, Germany}
\affiliation{L. D. Landau Institute for Theoretical Physics RAS, 
Kosygina street 2, 119334 Moscow, Russia}
\author{Alexey V. \surname{Ustinov}}
\affiliation{Physikalisches Institut,  
Karlsruhe Institute of Technology, D-76128 Karlsruhe, Germany}

\begin{abstract}
Experiments with one-dimensional arrays of Josephson junctions in the regime of dominating charging energy 
show that the Coulomb blockade is lifted at the threshold voltage, which is proportional to the array's length and 
depends strongly on the Josephson energy. We explain this behavior as de-pinning of the Cooper-pair-charge-density by the applied voltage. 
We assume strong charge disorder and argue that physics around the de-pinning point is governed by a disordered sine-Gordon-like model. This allows us to employ the well-known theory of charge density wave de-pinning. Our model is in good agreement with the experimental data. 
\end{abstract}

\date{\today}
\maketitle
One-dimensional Josephson arrays show a diverse range of transport regimes. 
In the regime of dominating Josephson energy, 
which attracts a continued experimental interest~\cite{Haviland2000,Ergul2013a,Ergul2013b}, they are highly conducting. In the regime of Josephson energy smaller or comparable to the charging energy, 
one-dimensional Josephson arrays show insulating (Coulomb blockade) behavior with activated transport~\cite{Zimmer2013}. 
Above a certain threshold value of the bias voltage, finite current appears even at zero temperature in the insulating regime. 
Initially, this switching was interpreted in terms of propagation onset of charge solitons~\cite{Ben-Jacob1989,Hermon1996,Haviland1996}, i.e., the 
energy one has to pay in order to push one soliton into the array. However, further experiments showed that the 
threshold voltage is proportional to the array length and depends strongly on the value of the 
Josephson energy~\cite{Haviland2000,Agren2001}. Here we interpret the experimentally found behavior as de-pinning in 
presence of strong charge disorder~\cite{Brazovskii2004}. 

We argue that the system is described by a model similar to a disordered sine-Gordon model. The only difference is the 
fact that, instead of the usual cosine potential, we have another periodic function, the lowest Bloch band energy, which 
depends strongly on the Josephson energy. It is this dependence which gives rise to the dependence of the switching voltage 
on the Josephson energy. Previously, similar models were derived~\cite{Hermon1996,Haviland1996,Haviland2000,Agren2001,Gurarie2004} using an additional phenomenological inductance in each cell of the array, which provided the necessary mass term. 
In Ref.~\onlinecite{Homfeld2011} it was shown, that a mass term is generated in the adiabatic regime due to the Bloch inductance~\cite{Zorin2006} and the phenomenological inductance is not needed. We argue that the adiabatic mechanism is sufficient 
to describe the system prior and at the de-pinning point.    

For this work, a series of experiments has been performed on a set of three
Josephson junction chains. The three arrays have been fabricated
in parallel on the same silicon substrate covered by an insulating thermally
grown SiO$_2$ layer. The individual cells of the array are
implemented as SQUID loops, similarly to earlier
experiments~\cite{Haviland1996,Haviland2000}.  The two tunnel junctions in each SQUID
are equivalent to a single junction with an effective Josephson energy $E_J(\Phi)$ tunable
by the magnetic flux $\Phi$ penetrating the loop area $A$. That gives
$E_J(\Phi)=E^m_J|\cos(\pi\Phi/\Phi_0)|$, where $E^m_J $ is twice the Josephson energy of one bare Josephson junction of the SQUID and $\Phi_0=h/2e$ is the magnetic flux quantum. 

The 
set of samples contains nominally identical arrays (labeled
A255, B255, and C255) comprising each 255 SQUIDs. These arrays
had very similar resistances.  Nevertheless, slight variations in the
junction parameters are reflected in the $I$-$V$
characteristics~\cite{Schafer2013}.  

\begin{figure}
\includegraphics[width=70mm]{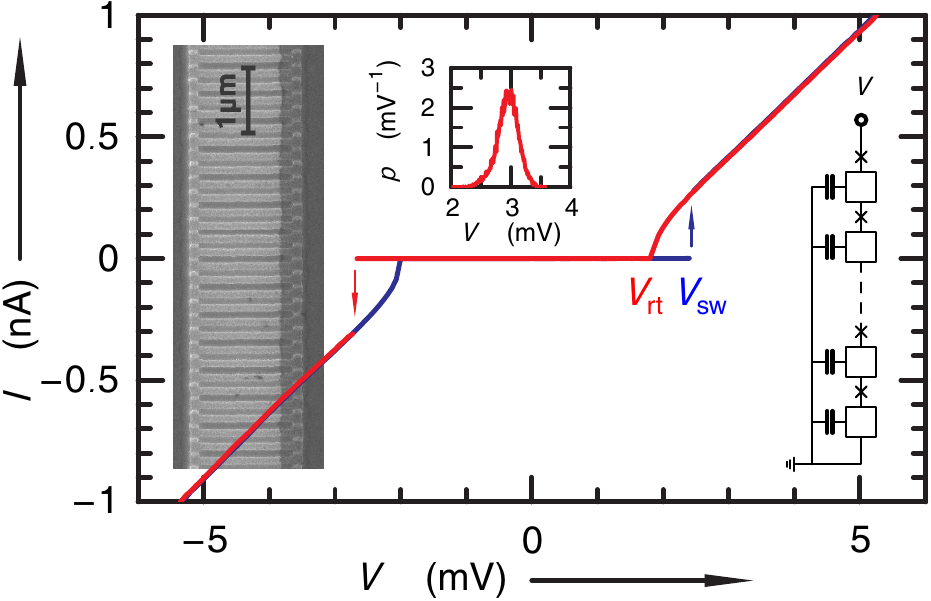}
\caption{\label{fig:i}
(color online) Hysteretic I-V characteristics of array
B255 measured for increasing voltages (blue) and decreasing
voltages (red) at $\Phi=0$.  The left inset shows an SEM micrograph of
the Josephson junction array; the right inset is a schematic
representation of the array. The middle inset displays the probability
density function of $V_\text{sw}$.}
\end{figure}

The experiments have been performed in a $^3$He/$^4$He dilution
refrigerator at 20\,mK temperature. A scanning electron microscope (SEM)
picture of a section of one of the arrays is shown in the left inset of
Fig.~\ref{fig:i}. All electrical connections to the samples are carefully
filtered by a combination of lumped-element low pass $RC$-filters and metal
powder filters covering a bandwidth of 10\,kHz. $I/V$ characteristics are
measured by ramping the applied bias voltage and recording the resulting
current with a homemade transimpedance amplifier. A typical $I/V$
characteristic is shown in Fig.~\ref{fig:i}, where the blue curve is
recorded while the bias voltage is ramped up and the red curve represents
the behavior for decreasing bias. In all cases, the current vanishes bellow
a certain threshold; for the horizontal branch, no current can be detected
within the resolution of our current measurement which is of the order of
50\,fA~\cite{Schafer2013}.  At a value $V_\text{sw}(\Phi)$, the chain switches to
a conducting state; the current after the switching is flux dependent and
has a magnitude of at least several pA. Retrapping to the $I=0$ branch
happens at a much lower voltage $V_\text{rt}<V_\text{sw}$. In this paper,
we focus on the magnitude and the flux dependence of the switching voltage
$V_\text{sw}(\Phi)$. We expect $V_\text{sw}$ to be primarily a function of
$E_J(\Phi)$. Thus, $V_\text{sw}(\Phi)$ is a periodic function in $\Phi$ with a
period of $\Phi_0$. Experimentally, we observe the period (measured in units
of the external magnetic field) to be of order $B_\text{ext}=6.9\,$mT,
corresponding to an area of $\Phi_0/6.9\,\text{mT}=0.3\text{\textmu{}m}^2$.
This agrees well with the total area per SQUID loop, $A_\text{SQUID}=
1.6\text{\textmu{}m}\cdot200\,$nm defined by the sample layout.

The rate by which the bias voltage at the sample can be changed is limited
by the bandwidth of the connecting leads (10\,kHz). In some cases we
recorded histograms for the switching voltage. The method used to
record switching histograms is detailed in Appendix~\ref{app:distribution}. A
typical example is shown as the middle inset in Fig.\ref{fig:i}. The
distribution of switching events turns out to be rather broad (e.\,g.\
$\sim1$\,mV for sample B255).
However, this measurements confirmed that the switching voltage as
extracted from single $I/V$ characteristics is close to the mean of the
histograms with a dispersion reflecting the width of the distribution.

The system is modeled as an array of superconducting islands 
(squares in the inset of Fig.~1) connected by Josephson junctions 
(crosses in the inset of Fig.~1). The junctions are characterized by the effective
Josephson energy $E_J$ (controlled by the magnetic field) and by the
effective capacitance $C_J\approx 2C_1$ ($C_1$ being the capacitance
of each of the SQUID junctions), 
which determines the 
(single electron) charging energy scale $E_C=e^2/2 C_J$.

Based on the area of the Al/AlO$_x$/Al tunnel junctions deduced from SEM micrographs we estimate that the average capacitance of the junctions is $C_J \approx 1 \,\text{fF} $. 
Due to variations in the areas of the tunnel junctions the values of $C_J$ are not necessarily constant along the array.

Screening, dominated by two ground planes running alongside of the array is modeled 
by attributing to each island a capacitance to the ground $C_0$ (see inset of Fig.~1).    
This gives the screening length $\Lambda\equiv\sqrt{C_J/C_0}$ and introduces 
yet another charging energy scale $E_{C0}\equiv e^2/2C_0=\Lambda^2 E_C$.
We estimate $5\,\text{aF}<C_0<20\,$aF.

In our theory we include disorder in the gate (frustration) charge $2e f_k$ 
on each superconducting island. 
The Hamiltonian then reads $H = H_C + H_J$, where 
\begin{equation}
\label{H_C}
	H_C = \frac{(2e)^2}{2}\, \sum_{k,q}  (n_k - f_k)  [C^{-1}]_{kq} (n_q - f_q)
\end{equation}	
and $H_J = -\sum_k E_J \cos\left({\theta}_k - {\theta}_{k+1} \right)$. Here $n_i$ is the number of Cooper-pairs on island $k$, 
and $[n_k,\exp(i\theta_q)]=\delta_{k,q}$.
The capacitance matrix is given by $C_{kq} =  \left(2 C_J + C_0 \right) \delta_{k,q} - C_J \left(\delta_{k-1,q} + \delta_{k+1,q} \right)$.
In the regime $C_J\gg C_0$, i.e., $\Lambda \gg 1$, one obtains $[C^{-1}]_{kq} \approx C_J^{-1}\,(\Lambda/2) \exp[-|k-q|/\Lambda]$.
Arrays with charge disorder in the limit $C_J \gg C_0$ have been considered 
long ago (see, e.g.,~\cite{Middleton1993,Johansson2000}). The onset of charge transport was calculated purely from 
the analysis of the stability of charge configurations. The crucial difference in our work is the strong renormalization 
of the disorder potential in the regime $E_J \sim E_C$.

The model introduced above was considered (without disorder) in Refs.~\onlinecite{Korshunov1989,RastelliPopHekking2013}
in the regime $E_J \gg E_C$ and also in Ref.~\onlinecite{Choi1998}. 
It was demonstrated that in the thermodynamic limit the system
undergoes a Beresinskii-Kosterlitz-Thouless quantum phase transition
and is an insulator for $K<2$, where $K\equiv \pi \sqrt{E_J/(8E_{C0})}=\pi\Lambda^{-1}\sqrt{E_J/(8E_{C})}$. Note, that due to $\Lambda \gg 1$ the regime 
$K\ll 1$ is compatible with $E_J \gg E_C$.

As realized in Refs.~\onlinecite{Korshunov1989,RastelliPopHekking2013}, in the regime $\Lambda \gg 1$ it is preferable 
to use the phase and change variables of the junctions rather than those of the islands. Thus we introduce 
the phase drops on the Josephson junctions $\phi_k \equiv \theta_{k}-\theta_{k+1}$
and their conjugate charge variables
$m_k \equiv  \sum_{p=1}^{k} n_p$. We express $H_C$ in terms of $m_k$ and after 
some algebra conclude that $H_C$ can be obtained by minimizing 
\begin{align}
	H_C\{Q\} &= \sum_{k=1}^{N} \left[\frac{\left(2e {m}_k - F_k - {Q}_k\right)^2}{2 C_J}  
	+\frac{\left({Q}_k-{Q}_{k+1}\right)^2}{2 C_0}\right]  \label{eqn:Ham_Q}
\end{align}
with respect to continuous charge variables $Q_k$. That is $H_C = min_Q[H_C\{Q\}]$. Here 
$F_k \equiv 2e \sum_{p=1}^{k} f_p$ is the accumulated random gate charge. 
The quasi-charges $Q_k$ are well known in the theory of Coulomb blockade and appear naturally in the 
theories including a phenomenological inductance~\cite{Hermon1996,Gurarie2004}. Their electrostatic meaning 
and the derivation with inductances is explained in Appendix~\ref{app:inductances}. 

The introduction of $Q_k$ is equivalent to a Hubbard-Stratonovich transformation in the sense that, e.g., 
the real time (Keldysh) partition function can be obtained as
$
Z ={\cal N}\int \prod_k DQ_k D m_k D \phi_k \, e^{i \int dt \left[\sum_k  m_k \dot \phi_k - H_C\{Q\} - H_J\right]}
$, 
where ${\cal N}$ is a normalization factor. For a given path of the quasi-charges $Q_k(t)$ the $(m_k,\phi_k)$-dependent part of the Hamiltonian $H_C\{Q\} + H_J$ separates into Hamiltonians of independent Josephson junctions biased each by charge  $Q_k +  F_k$, i.e., 
\begin{align}\label{Hi}
	H_k &=  \frac{1}{2 C_J} \left(2e {m}_k - {Q}_k -  F_k \right)^2 - E_J \cos{\phi}_k\ .
\end{align}
To obtain the effective quasi-charge theory we integrate out the discrete charge degrees of freedom
$m_k, \phi_k$. At temperatures much lower than the band gap of (\ref{Hi}), i.e., the $Q$-dependent 
energy splitting between the ground and the first excited states of (\ref{Hi}), and close enough to equilibrium 
it should be sufficient~\cite{Gurarie2004} to consider only adiabatic paths $Q_k(t)$ as was done in
Ref.~\onlinecite{Homfeld2011}. These are paths that do not induce 
Landau-Zener transitions between the energy bands of (\ref{Hi}). 
Generalizing the derivation of Ref.~\onlinecite{Homfeld2011} to the regime of charge disorder and defining $Q^F_k \equiv Q_k + F_k$
we obtain the following effective Lagrangian
\begin{align}
	{\cal L} &= \sum_{k} \left[\frac{L_B(Q^F_k)\,\dot{Q}_k^2}{2}  - \frac{ \left(Q_k-Q_{k+1}\right)^2 }{2 C_0}
	- U \left[Q^F_k\right] \right] . \label{eqn:QLagrangian} 
\end{align}
Here $L_B(Q)$ is the Bloch inductance~\cite{Zorin2006,Homfeld2011} whereas 
$U\left[Q\right]$ is the zeroth Bloch band energy ($Q$-dependent ground state energy of (\ref{Hi})). 
Thus, the mass term $\propto L_B$ is generated and the phenomenological inductance used in~\cite{Hermon1996,Gurarie2004} 
is not necessary. In this paper, we are interested in depinning and approach this transition from the non-dynamical pinned side, where fast changes in the quasi-charge are naturally suppressed. Thus, we argue that the description in terms of slow adiabatic paths $Q_k(t)$ is applicable, at least for not very small values of $E_J$. This assumption will be checked for self-consistency below. 

Our central idea here is that in the regime $E_J \sim E_C$ and $\Lambda \gg 1$ the model 
(\ref{eqn:QLagrangian}) is still applicable whereas the pinning potential is strong and varies significantly with varying 
$E_J(\Phi)$. This explains the strong dependence of the switching voltage on $\Phi$. 
The idea of classical charge pinning in Josephson arrays was first proposed by Gurarie and Tsvelik~\cite{Gurarie2004}.
There, the classical regime $K \ll 1$ was achieved by introducing a phenomenological large inductance~\cite{Hermon1996}. 
Our main achievement here is in showing that Bloch inductance is sufficient to render the pinning regime. 

To describe the onset of transport (depinning) it is sufficient to focus on the potential energy 
part of (\ref{eqn:QLagrangian}). In the continuum limit justified by large $\Lambda$, we obtain the following well 
established continuum model for CDW-depinning~\cite{Brazovskii2004} 
\begin{align}
	H_C &= \int dx\left[ \frac{(\partial_x Q(x))^2}{2 C_0}  + U\left[Q(x)+ F(x)\right]
	- E\, Q(x)\right] \ , \label{eqn:Ham_CDW}
\end{align}
where the spatial coordinate $x$ is measured in units of the array lattice constant. Here
$E \equiv V/N$ is the homogeneous depinning force (electric field). In Appendix~\ref{app:bias} 
we discuss the case of the bias voltage applied at the edge. 

We assume a strong (maximal) charge disorder, i.e., the gate charges 
$2e f_k$ being homogeneously distributed in an interval of length $2e$ or larger.
This is equivalent to a homogeneous distribution of $F_k$ between $-e$ and $e$ and
statistical independence of $F_k$ and $F_q$ for $k\neq q$. Indeed, the disorder charge $F_k$ is effectively 
limited to the interval $\left[-e,e\right]$ as any deviation thereof is compensated by adjusting 
the number of Cooper-pairs on the islands. 

As discussed in detail in the literature (for review see Ref.~\onlinecite{Brazovskii2004}), the critical value of the depinning force is determined by the competition between the disorder pinning potential and the elastic energy. The two become comparable 
at the so called Larkin length $N_L$ (a.k.a. Fukuyama-Lee or Imry-Ma length)  \cite{Larkin1970,Imry1975,Fukuyama1978} and at 
$E_p \approx e(C_0 N_L^2)^{-1}$ the charge is depinned
\footnote{$E_p$ corresponds here to the depinning force $f_p$ in the notations of Ref.~\onlinecite{Brazovskii2004}. This estimate can be improved with the help of an RG treatment~\cite{Brazovskii2004, Chauve2000a, Chauve2001}.
The improved estimate is called $f_c$ in Ref.~\onlinecite{Brazovskii2004}.  The relevance of this RG procedure 
in our case is questionable since our arrays are not much longer than Larkin length}.

The Larkin length is calculated ~\cite{Fukuyama1978} using the pinning strength $R$ of the effective potential $U(Q)$, 
\begin{align}
  R =& \textrm{max}_{Q \in [-e,e]}\left[U(Q)\right]-\textrm{min}_{Q \in [-e,e]}\left[U(Q)\right]\ .
\end{align} 
One obtains for the Larkin length~\cite{Fukuyama1978} 
\begin{align}
  N_L    	\approx 3^{-2/3} \Lambda^{4/3} \tilde{R}^{-1/3}\ ,\label{eqn:Larkin_length} 
\end{align}
where $\tilde{R}\left[E_J(\Phi)/E_C\right]	 \equiv  \frac{1}{16 E_C^2}  R^2 $.
The dependence of $\tilde{R}$ on the dimensionless parameter $E_J(\Phi)/E_C$ is obtained numerically (see Appendix~\ref{app:corr_fun}).

Thus, we obtain the following estimate for the switching voltage
\begin{align}
  V_{sw} & =  N E_p \approx \frac{2 N E_C}{e} 3^{\frac{4}{3}} \, \Lambda^{-\frac{2}{3} }  \tilde R^{2/3}  \ . \label{eqn:sw_Voltage} 
\end{align}
This expression is valid as long as the Larkin length is much shorter than the array length, $N_L \ll N$.

\begin{figure}
\includegraphics{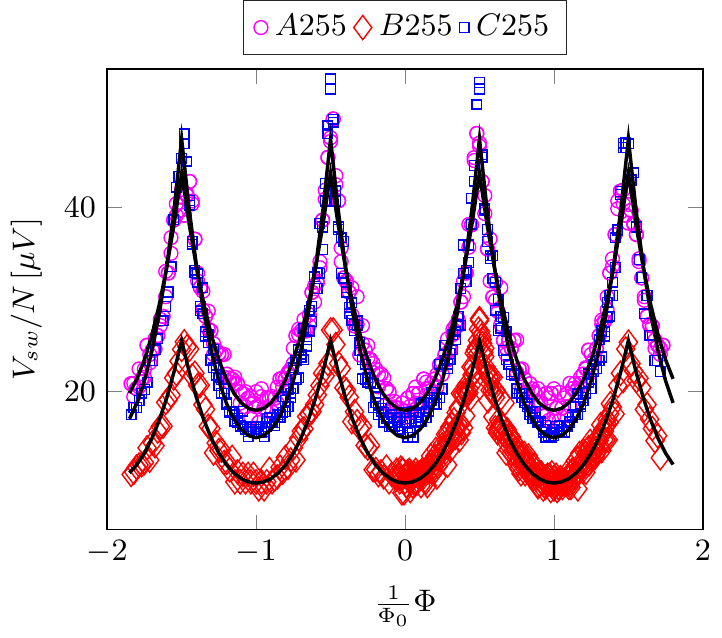}
\caption{ \label{fig:fit_AtoC} (color online) The switching voltage normalized to the array length N as a function of the magnetic flux $\Phi$ for three arrays of length $255$. Solid lines are fitted functions, circles show experimental data. }
\end{figure}

We use Eq.~(\ref{eqn:sw_Voltage}) to fit the experimental data for arrays $A255$, $B255$, and $C255$ 
(see Fig.~\ref{fig:fit_AtoC}). From the device fabrication one can expect the value of the ground capacitance $C_0$ to vary only to a small degree between the samples. At the same time an exact value for $C_0$ can not be determined experimentally. The other parameters of the array islands, $C_J$ and $E_J^m$, can vary between different samples and also between different islands of the same sample due to imperfections in the junctions. We use the obtained fitting parameters to express the effective $C_J$ and $E_J^m$ in terms of the undetermined $C_0$ and give the values corresponding to either $C_0 = 5 \textrm{aF}$ or $C_0 = 20 \textrm{aF}$ (Tab.~\ref{tab:fitting_values}). We obtain values of $C_J$ and $E_J^m$ that are comparable with the ones expected from geometrical estimates.
(Given the uncertainty of the numerical coefficients in (\ref{eqn:sw_Voltage}), some deviations should be expected.)
As the Larkin length $N_L$ depends on $E_J$, we only provide the maximal value $N_L^{\rm max}$, achieved 
at $\Phi=0$, where $E_J^{\phantom m} = E_J^m$, and the minimal value $N_L^{\rm min}$, achieved 
at $\Phi=\Phi_0/2$, where $E_J \approx 0$. The depinning approach is applicable since $N_L < N$.

\begin {table}[H]
\begin{center}

\begin{tabular}{|c|c|c|c|}
	\hline
array 	 & 	A255	& B255	 & C255 \\ \hline
 $N$ 	 &	$255$  	& $255$  	&  $255$	  	\\ \hline
 $C_J\Lambda^{\frac{2}{3}} = C_J^{\frac{4}{3}}C_0^{-\frac{1}{3}} $	 & 	$  2.5 \pm 0.01 \, \textrm{fF} $  	& $ 4.27 \pm 0.03 \, \textrm{fF} $ &  $  2.3 \pm 0.01 \, \textrm{fF}  $   	 	\\ \hline
$E_J^m/E_{C}$ & $ 1.27 \pm 0.02 $ &  $ 1.33 \pm 0.02$ &  $ 1.63 \pm 0.02$	\\ \hline
$C_{J\,\,(C_0 \approx 5 \textrm{aF})}$ & $0.53$fF &  $0.79$fF &  $0.5$fF \\ \hline
$\Lambda_{\,\,(C_0 \approx 5 \textrm{aF})}$ & $10.3$ &  $12.6$ &  $10$ \\ \hline
$E^m_{J\,\,(C_0 \approx 5 \textrm{aF})}$ & $192 \mu$eV & $134 \mu$eV & $ 262 \mu$eV  \\ \hline
$N^{\rm min/max}_{L\,\,(C_0 \approx 5 \textrm{aF})}$ & $[27,42]$ & $[35,56]$ & $[26,46]$  \\ \hline
$C_{J\,\,(C_0 \approx 20 \textrm{aF})}$ & $0.75$fF &  $1.12$fF &  $0.7$fF \\ \hline
$\Lambda_{\,\,(C_0 \approx 20 \textrm{aF})}$ & $6.1$ &  $7.5$ &  $6.0$ \\ \hline
$E^m_{J\,\,(C_0 \approx 20 \textrm{aF})}$ & $136 \mu$eV & $95 \mu$eV & $ 186 \mu$eV  \\ \hline
$N^{\rm min/max}_{L\,\,(C_0 \approx 20 \textrm{aF})}$ & $[13,21]$ & $[18,28]$ & $[13,23]$  \\ \hline

\end{tabular}

\caption{\label{tab:fitting_values} The experimental estimates and fitted values for Josephson junction arrays 
A255, B255, and C255.}
\end{center}
\end {table}

When comparing to other previously explored models we notice the difference between the physics we describe here and the de-pinning of a single charge soliton in a disordered array.  The latter case was analyzed within the disordered sine-Gordon model~\cite{Fedorov2011}. It was shown that   the depinning critical force grows with the soliton length $\Lambda$. In our case, however, the depinning transition is a collective phenomenon in the whole array. At the transition point the array contains, on average, one extra charge of $2e$ per Larkin length, $N_L \propto \Lambda^{4/3} \tilde R^{-1/3}$. The longer is $\Lambda$, the fewer charges are pinned and the easier is the depinning, $E_p \propto \Lambda^{-\frac{2}{3}} \tilde R^{2/3}$. As mentioned above, models of transport onset that rely on the creation of a propagating soliton~\cite{Haviland1996} can not explain the linear dependence of $V_{sw}$ on $N$, observed in experiments.

We, finally, check the consistency of our adiabatic assumption. Clearly, it is well justified if $E_J \gg E_C$ and it must break down 
if $E_J \ll E_C$. To get a more precise criterium, we assume $E_J \approx E_C$ and estimate the typical oscillation (pinning) frequency of a domain of length $N_L$ with a rigid quasi-charge $Q$. We obtain $\omega_{p}\sim \sqrt{\frac{2 E_J E_C}{\sqrt{N_L}}}$ (cf.~\cite{Feigelman1980,Aleiner1994,Fogler2002,Gurarie2004}). We compare this 
with the plasma frequency $\sqrt{8E_J E_C}$, which is this regime is also of the order of the critical Landau-Zener frequency. We conclude  
that, parametrically, for $N_L \rightarrow \infty$, e.g., for $\Lambda \rightarrow \infty$, the adiabatic assumption is well justified. More precise estimates show that, for our arrays, the adiabatic assumption is valid except for a narrow domain of $\Phi$ around $\Phi_0/2$ where 
$E_J(\Phi)\ll E_C$.         

In this paper we have compared the experimentally measured magnetic flux dependence of the switching voltage of an insulating (Coulomb blockaded) SQUID-array with our theoretical predictions based on a sine-Gordon-like model for a continuous quasi-charge field. Based on Ref.~\onlinecite{Homfeld2011} we argue that this model can be applied without introducing artificial large inductances~\cite{Haviland1996,Haviland2000, Hermon1996,Gurarie2004}. We employ the 
connection to the theory of charge density wave depinning, first pointed out in Ref.~\onlinecite{Gurarie2004}, to theoretically analyze the switching voltage and fit the experimental data. We find that the breakdown of the insulating state in Josephson junction arrays is a collective depinning effect, similar to that of depinning of charge density waves, vortices in type II superconductors etc. The switching behavior of Josephson junction arrays can therefore be linked to a rich research area of physics. We think this could be particularly interesting as Josephson junction arrays are artificially fabricated and could possibly help us to study depinning physics in the limit of very short systems or at the crossover from discrete systems to the continuum limit. Transport well above the switching voltage remains the subject of continuing investigations~\cite{Cole2014}. It  will be interesting to match this transport regime with the depinning physics analyzed in this paper.
\begin{acknowledgments}
We thank A.D. Mirlin, J.H. Cole, B. Kie{\ss}ig,  D.G. Polyakov, S.V. Syzranov, and T. Giamarchi for helpful discussions of the subject. The theory analysis was funded by the Russian Science Foundation under Grant No. 14-42-00044.
\end{acknowledgments}

\appendix

\section{Array with inductances}
\label{app:inductances}

\begin{figure}
\includegraphics[width=0.9\columnwidth]{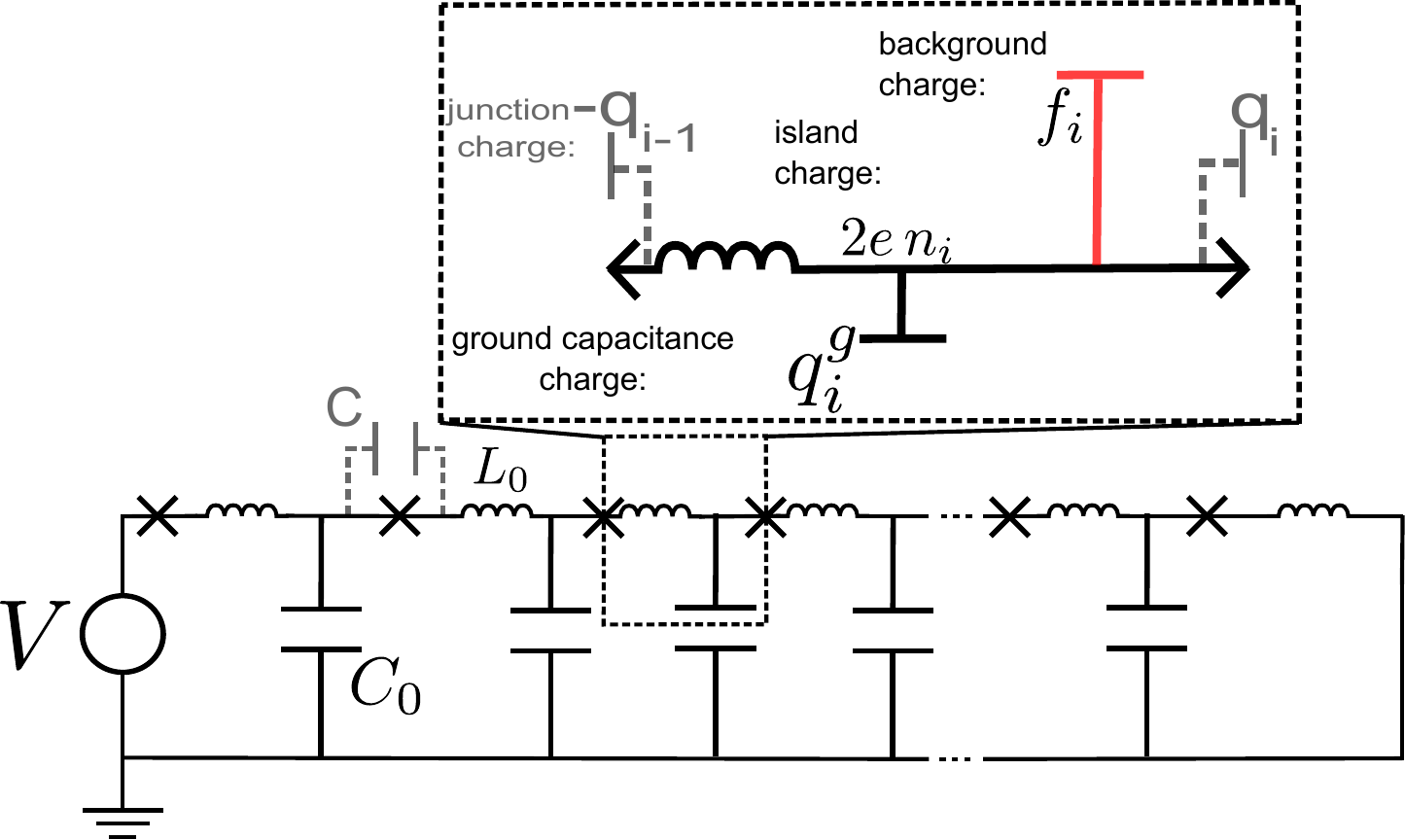}
\caption{ \label{fig:arraywithL} (color online) Sketch of the Josephson junction array with inductances $L_0$. 
The magnified part shows the distribution of the charges $q_i^{g}$, $q_i$, $2 n_i$ and $f_i$ on the island and the capacitances. In the language of electrical circuits the background charge $f_i$ is given by the constant charge on an additional capacitance that is connected to the island, as shown in red in the magnified sketch of the island above. }
\end{figure}

The introduction of inductances $L_0$ into the model (see Fig.~\ref{fig:arraywithL}) 
necessitates a description in terms of continuous as well as discrete charge variables. The discrete ones are  
the overall charges $2e n_i$ of the islands. The continuous ones are the charges $q_i$ on the junction 
capacitances $C_J$ and charges $q^{g}_{i}$ on the capacitances to the ground $C_0$. Conservation of charge requires 
\begin{align}
	2e\,n_i - f_i - q_i^{g} + q_{i-1} - q_{i} = 0\ , \label{eqn:AppChargeConserv}
\end{align}
where $f_i$ are the random offset charges. 
Introducing the integrated charge variables 
$m_i \equiv  \sum_{j=1}^{i} n_j$,  $Q_i = 2e \sum_{j=1}^{i} q^{g}_j$, and $F_i = 2e \sum_{j=1}^{i} f_j$ one can easily obtain the following hamiltonian
	\begin{eqnarray}\label{HwL}
			H&=& \sum_{i=1}^{N}\left[ \frac{1}{2 C_J} \left(2e {m}_i -  F_i - {Q}_i\right)^2 - E_J \cos\phi_i\right.\nonumber 
			\\&+&\left. \frac{1}{2 C_0} \left({Q}_i-{Q}_{i+1}\right)^2 + \frac{1}{2 L_0} \Phi^2_i \right]\ ,
	\end{eqnarray}
where $\Phi_i$ is the flux on the inductance $L_0$ of the i-th island whereas $\phi_i$ is the phase drop 
on the i-th Josephson junction. The pairs of canonically conjugated 
variables in (\ref{HwL}) are $(Q_i,\Phi_i)$ and $(m_i,\phi_i)$.
The physical meaning of $Q_i$ is clarified by the following relation
\begin{align}
		q_i &= Q_i+ F_i - 2e m_i\ ,
\end{align}
which can be obtained using (\ref{eqn:AppChargeConserv}). The charge on the junction capacitance $q_i$ is 
given by the total charge that has flown into the junction $Q_i + F_i$ minus the discrete charge $2e m_i$ that has tunneled 
through the junction. As $F_i$ is a constant offset charge, we understand that $Q_i$ is the integral of current that has flown 
into the junction.

In Ref.~\onlinecite{Hermon1996} the inductance $L_0$ was assumed to be large, so that the 
dynamics of $(Q_i,\Phi_i)$ is adiabatic. In the current paper we assume $L_0 \rightarrow 0$ and 
claim that the emerging Bloch inductance, the large screening length $\Lambda$ and the pinning disorder
render an adiabatic regime in the vicinity of the depinning point. 

\section{Relation to Luttinger liquid}
\label{app:LL}

In the limit $E_J \gg E_C$ the Bloch inductance $L_B$ approaches~\cite{Zorin2006,Homfeld2011}  
the Josephson inductance $L_J \equiv (\Phi_0/(2\pi))^2 E_J^{-1}$, whereas $U[Q] \approx -E_S\cos\left[2\pi Q/(2e)\right]$. 
Here $E_S$ is the exponentially small phase slip amplitude~\cite{Korshunov1989,RastelliPopHekking2013}.
Introducing $q_k = \pi Q_k/(2e)$ we obtain from Eq.~(\ref{eqn:QLagrangian}) the discretized Lagrangian of the
Luttinger liquid~\cite{giamarchi2003quantum} with phase disorder in the backscattering term
\begin{align}
	{\cal L} &=\frac{1}{2\pi K} \sum_{k} \left[\frac{\dot{q}_k^2}{v} - v \left(q_k-q_{k+1}\right)^2\right] \nonumber\\
	&+\sum_k E_S\cos\left[2q_k +\pi F_k/e\right] \,. \label{eqn:LuttingerLagrangian} 
\end{align}
Here $v\equiv 1/\sqrt{L_J C_0}$ and $K\equiv \pi \sqrt{E_J/(8E_{C0})}$. Thus, for $F_k=0$, i.e., without disorder, 
we reproduce the conclusions of Refs.~\onlinecite{Korshunov1989,RastelliPopHekking2013}. In the limit 
$\Lambda \rightarrow \infty$ we obtain $K\rightarrow 0$ and the relevant physics in the thermodynamic limit is 
that of classical pinning~\cite{giamarchi2003quantum,Mirlin2007Pinning}. Yet, since in the limit $E_J \gg E_C$ the 
pinning potential $\sim E_S$ is exponentially weak, systems of finite length may conduct or even be superconducting~\cite{RastelliPopHekking2013}.

\section{Voltage bias at the edge}
\label{app:bias}

We consider the potential energy part of the Hamiltonian of the Josephson junction array,
\begin{align}
H_C &= \sum_i \,\left[ \frac{\left(Q_i-Q_{i+1}\right)^2 }{2 C_0}  + U \left[Q_i +  F_i\right]  \right] 
- Q_{i=1} V .
\end{align}
Here the last term has been added to describe the voltage bias $V$ applied on the left edge of the array. 
To transform an edge voltage bias to a homogeneous electric field we perform the following transformations 
$\tilde Q_i \equiv Q_i -  A_i$ and  $\tilde F_i  \equiv   F_i  + A_i$, where $A_i\equiv  C_0 V (N+1-i)(N-i)/2N$
and $N$ is the length of the array. This gives
\begin{align}
H_C &= \sum_i \,\left[ \frac{ \left(\tilde Q_i-\tilde Q_{i+1}\right)^2}{2 C_0}  + U \left[\tilde Q_i +  \tilde F_i\right]   -  
E\,\tilde Q_i \right]\  ,
\end{align}
where $E \equiv V/N$ is the homogeneous depinning force (electric field).
In the case of maximal disorder the shift of the quasi-charge to include the voltages applied 
at the boundaries does not change the distribution function of the random charge $\tilde F_i$. 
Thus we can omit the tildes and we obtain the model of Eq.\ref{eqn:Ham_CDW}. This property of the maximally disordered 
model is also referred to as statistic tilt symmetry~\cite{Fedorenko2006}. 

\section{The strength of the pinning potential}
\label{app:corr_fun}

\begin{figure}
\tikzsetnextfilename{R_tilde_plot}
\includegraphics{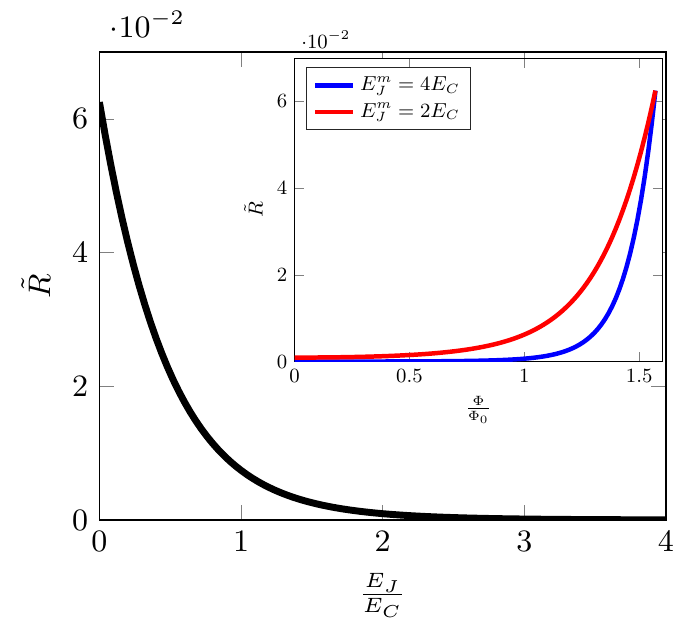}
\caption{ \label{fig:R_tilde} (color online) The dimensionless strength of the pinning potential $\tilde{R}$ as a function of $E_J/E_C$ in the main plot and as a function of the magnetic flux $\Phi$ in the inset plot.
}
\end{figure}

The strength of the pinning potential $R$ can be obtained by numerically diagonalizing the single junction Hamiltonian 
\begin{eqnarray}
	H(Q) = 4 E_C\left( \left( \hat{m} -\frac{Q}{2e}\right)^2 + \frac{E_J}{8 E_C} \left( \left| m+1 \right\rangle \left\langle m \right|+ \textrm{h.c.} \right) \right) \nonumber \\
\end{eqnarray}
for a dense set of $Q$-values in the interval $[-e,e]$. For diagonalisation we use $15$ charge state $\left|m\right\rangle$ with
lowest charging energy. Including more states does not change the ground state energy $E_Q(Q)$ within our level of numerical accuracy. The value of the function $\tilde{R}$ for each fixed value of $E_J/E_C$ can be obtained by determining the amplitude of the periodic function $E_Q(Q)$. The result is shown in Fig.~\ref{fig:R_tilde}.
\tikzexternalenable

\section{Switching voltage as a function of Josephson coupling energy}
\begin{figure}
\tikzsetnextfilename{A_to_C_linear}
\includegraphics{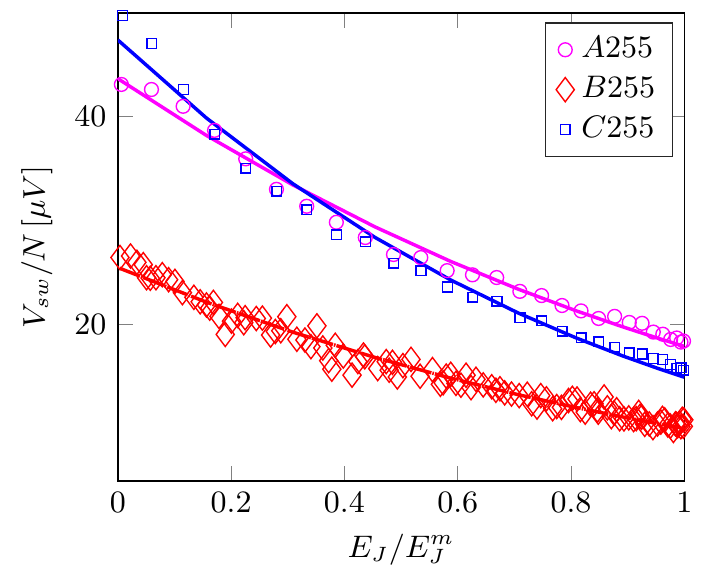}
\caption{ \label{fig:A_to_C_linear}(color online) The switching voltage as a function of $E_J$ for arrays $A255$ (magenta), $B255$ (red) and $C255$ (blue). }
\end{figure}

One of the dominant effects visible in Fig.~\ref{fig:fit_AtoC} is the periodicity of the switching voltage $V_{sw}$ with magnetic flux. This periodicity is a consequence of the periodicity of the Josephson coupling energy $E_J \propto \cos(\frac{\pi\Phi}{\Phi_0})$. The switching voltage $V_{sw}$ is plotted as a function of $E_J$ in 
Fig.~\ref{fig:A_to_C_linear}.

\tikzexternalenable
\section{Distribution of the switching voltage}
\label{app:distribution}
\begin{figure}

\includegraphics[width=70mm]{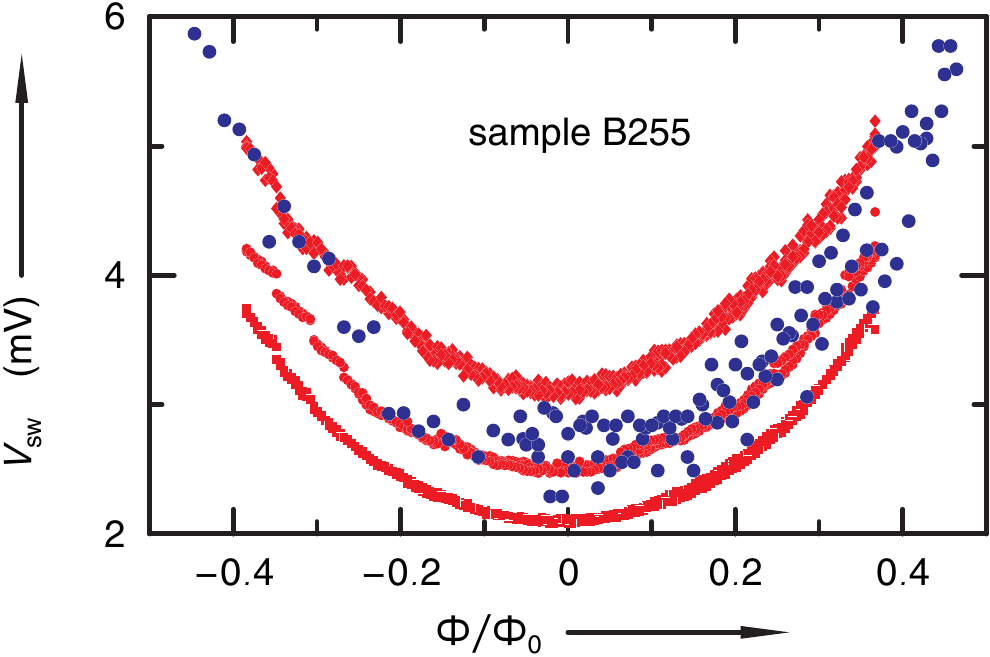}

\caption{\label{fig:siii}(color online) Comparison of switching voltage extracted from
single sweeps (blue dots) and full switching voltage histograms (red
symbols). The data displayed in red show the width of the histograms (see
text for an explanation).}

\end{figure}

The switching voltage $V_\text{sw}$ shows strong fluctuations. The data
presented in Fig.~\ref{fig:fit_AtoC} are extracted from individual
measurements of $I/V$ characteristics. The bias voltage was ramped up once
and the current response was detected by a transimpedance amplifier. In
these measurement switching can easily be identified as evident from
the sample IV-curve given in Fig.~\ref{fig:i}.  The fluctuation in $V_\text{sw}$ can be noticed from the
apparent noise visible in Fig.~\ref{fig:fit_AtoC}. For the sample B255 we have 
recorded many switching
events at various fixed values of $\Phi$ and constructed histograms.  The result of these measurements is
summarized in Fig.~\ref{fig:siii}, where the properties of histograms are
visualized by red symbols and single switching events extracted from
individual $I/V$ characteristics are shown as blue dots. The latter data are
the same as displayed in Fig.~\ref{fig:fit_AtoC}. Histograms
are constructed from at least 10000 switching events. The events are sorted
according to its switching voltage $V_\text{sw}$ and the range of
$V_\text{sw}$ is divided in about 250 to 300 bins. To construct histograms,
the events corresponding to each bin are counted.  The mean $V_\text{mean}$
of the histograms (this is 50\,\% of times the switching occurs at voltages
lower than $V_\text{mean}$) is represented as red dots in
Fig.~\ref{fig:siii}.  The red squares correspond to the voltage of the
lowest bin containing at least 0.15\,\% of the events, the red diamonds to
the highest bin containing at most 0.15\,\% of the event. The vertical
distance between the squares and diamonds represent thus the full width of
the histograms.  Single events as seen in $I/V$ characteristics (blue dots)
fall well into the span of switching voltages recorded in a quite different
manner for the purpose to construct the histograms.

The method to record a great number of events is 
 rather conventional. A sawtooth like voltage signal with $0<V<V_\text{max}$
has been applied as bias to sample B255 where $V_\text{max}$ is
considerable larger than the maximally observed switching voltage. Each
time the bias starts to ramp at $V=0$ a timer is started. The voltage
output $V_o$ of a transimpedance amplifier is used as a trigger signal to
stop the timer as soon as $V_o$ exceeds a threshold signaling that
switching from a zero current to a finite current state has occured.
Retrapping occurs when the bias is set back to zero at the end of each
voltage ramp. The time span between start and stop trigger is a measure of
the switching voltage of a single event. The frequency of the sawtooth
signal is of the order of 20\,Hz and the recording of a histogram takes
about 10 minutes.

The current needs to be detected with a
relative large bandwidth. The resolution of the current measurement is for
this reason considerable worse than the resolution achieved in measurements
of individual $I/V$ characteristics. In the latter case the bias voltage
can be varied very slowly while the output of the transimpedance amplifier is
averaged to yield the desired current resolution. To construct a histogram
many events have to be measured and a histogram can be constructed in a
reasonable time only when the current after switching is sufficiently large
to be detected quickly. Since the current after switching is getting
smaller close to full frustration $\Phi=(n+1/2)\Phi_0$ histograms could
only be measured in the range of small frustrations dipicted in
Fig.~\ref{fig:siii}

\bibliography{Promotion-depinning_paper}

\end{document}